\documentclass[12pt]{iopart}

\usepackage{iopams} 
\expandafter\let\csname equation*\endcsname\relax

\expandafter\let\csname endequation*\endcsname\relax
\usepackage{amsmath}
 
\usepackage{trfsigns}

\usepackage{graphicx}
\usepackage{siunitx}
\usepackage{amssymb}
\usepackage{color}
\usepackage[latin1]{inputenc}
\usepackage{mathrsfs}
\usepackage{subfigure}
\usepackage{cite}
\usepackage{booktabs}
\renewcommand{\vec}[1]{\boldsymbol{#1}}

\usepackage{wasysym}
\usepackage[hidelinks]{hyperref}
\newcommand{\bichrom}[2]{($#1\omega$:$#2\omega$)}
\renewcommand{\d}[1]{\ensuremath{\operatorname{d}\!{#1}}}
\usepackage{soul}

\begin{document}
	
	\title[Orbital angular momentum superposition states in TEM and MPI]{
		Orbital angular momentum superposition states in transmission electron microscopy and bichromatic multiphoton ionization
	}
	\author{K. Eickhoff$^1$\footnote[7]{These authors contributed equally to this work.}, C. Rathje$^2$\footnotemark[7], D. K\"ohnke$^1$, S. Kerbstadt$^{1,3}$,\\
		L. Englert$^1$, T. Bayer$^1$, S. Sch\"afer$^2$, M. Wollenhaupt$^1$}
	\address{$^1$ Carl von Ossietzky Universit\"at Oldenburg, Institut f\"ur Physik, Ultrafast Coherent Dynamics, Carl-von-Ossietzky-Stra\ss e 9-11, D-26129 Oldenburg, Germany}
	\address{$^2$ Carl von Ossietzky Universit\"at Oldenburg, Institut f\"ur Physik, Ultrafast Nanoscale Dynamics, Carl-von-Ossietzky-Stra\ss e 9-11, D-26129 Oldenburg, Germany}
	\address{$^3$ Center for Free-Electron Laser Science (CFEL), Deutsches Elektronen Synchrotron DESY, Notkestraße 85, D-22607 Hamburg, Germany}
	
	\ead{Sascha.Schaefer@uol.de, Matthias.Wollenhaupt@uol.de}
	\vspace{10pt}
	
	\begin{abstract}
		The coherent control of electron beams and ultrafast electron wave packets dynamics have attracted significant attention in electron microscopy as well as in atomic physics. In order to unify the conceptual pictures developed in both fields, we demonstrate the generation and manipulation of tailored electron orbital angular momentum (OAM) superposition states either by employing customized holographic diffraction masks in a transmission electron microscope or by atomic multiphoton ionization utilizing pulse-shaper generated carrier-envelope phase stable bichromatic ultrashort laser pulses. Both techniques follow similar physical mechanisms based on Fourier synthesis of quantum mechanical superposition states allowing the preparation of a broad set of electron states with uncommon symmetries. We describe both approaches in a unified picture based on an \emph{advanced spatial} and \emph{spectral double slit} and point out important analogies. In addition, we analyze the topological charge and discuss the control mechanisms of the free-electron OAM superposition states. Their generation and manipulation by phase tailoring in transmission electron microscopy and atomic multiphoton ionization is illustrated on a 7-fold rotationally symmetric electron density distribution.

	\end{abstract}
	%
	%
	%
	%
	%
	
	\section{Introduction} 
	Spatially coherent electron probes have developed into a versatile tool for exploring the quantum nature of matter on the atomic scale \cite{Muller.2009}. Especially electron orbital angular momentum (OAM) beams with tailored symmetries and topologies \cite{Bliokh.2007,Bliokh:2017:PR:1,Lloyd.2017} opened up a new degree of freedom in quantum control scenarios and may provide selective access to additional material properties. Recently, these beams have enabled quantized OAM transfer to atoms in electron-energy loss spectroscopy (EELS) \cite{Verbeeck:2010:Nature:301,Lloyd:2012:PRL:074802} and were proposed for the characterization of chiral crystal symmetries in electron diffraction \cite{Juchtmans:2015:PRB:134108}. Inducing magnetic transitions in atoms using OAM electron beams may provide a tool to probe magnetic states of matter on the nanoscale \cite{Schattschneider:2014:U:81}. Furthermore, electron beams with unusual topology exhibit intricate field-interaction effects, including free-electron Landau states \cite{Schattschneider.2014b}.\\
	Optical OAM beams \cite{Allen:1992:PRA:8185,Torres:2011,Shen:2019:LSA:1} already found a variety of applications ranging from fundamental physics \cite{Bezuhanov:2004:OL:1942,Surzhykov:2015:PRA:013403,Matula:2013:JPB:205002,Wang:2018:N:1533} to optical tweezers \cite{Liesener:2000:OC:77,ONeil.2002} and spanners \cite{Simpson:1997:OL:52}.
	In contrast to the micrometer-sized foci of optical OAM beams, for OAM electron beams the orbital angular momentum plays a more significant role in electron-matter interaction, due to their nanometer-sized foci overlaping with atomic-scale quantum systems \cite{Verbeeck.2011,Guzzinati.2017}. Therefore, electron OAM states, characterized by a phase singularity, a helical phase front and a non-vanishing topological charge~$\ell$ \cite{Bliokh.2007,Bliokh:2017:PR:1,Lloyd.2017}, are the subject of current research.\\
	In recent years, several experimental techniques for the generation of high-quality OAM electron beams were developed for transmission electron microscopy (TEM), including the application of diffraction holograms \cite{Verbeeck:2010:Nature:301,McMorran.2011} and phase masks \cite{Uchida:2010:Nature:737,Shiloh.2014,Grillo.2014,Beche.2016,Grillo.2017}, as well as by using the quasi-monopole magnetic field of a thin magnetic needle \cite{Beche:2014:NP:26}. Furthermore, inelastic \cite{Barwick.2009,Park.2010,GarciadeAbajo.2010b,Piazza.2015,Feist.2015,Priebe.2017,Vanacore.2018,Feist.2020,Harvey.2020} and elastic \cite{Kapitza.1933,Freimund.2001,Schwartz.2019} electron-light scattering have been demonstrated to facilitate a detailed control of the phase structure of electron beams.\\
	In ultrafast and attosecond spectroscopy of atoms, coherent control of photoemitted electrons was utilized to obtain a detailed picture of light-driven strong-field ionization channels, including perturbative \cite{Shapiro:2003} and non-perturbative multiphoton ionization (MPI) \cite{Brabec:2008} and tunneling processes \cite{Vrakking:2014:PCCP:2775,Krausz:2009:RMP:163,Corkum:2007:NP:381}.  
	Polarization-tailored bichromatic \bichrom{n}{m} fields \cite{Eichmann:1995:PRA:R3414,Milosevic:2000:PRA:063403,Fleischer:2014:NP:543,Ehlotzky:2001:PR:176,Kerbstadt:2017:OE:12518,Kerbstadt:2017:JMO:1010} have been established as powerful tools in coherent quantum control \cite{Kfir:2015:NP:99} and were shown to address optically controlled quantum interferences between pre-defined electron wave functions \cite{Kerbstadt:2018:PRA:063402,Kerbstadt:2019:NC:658,Ehlotzky:2001:PR:176,Kerbstadt:2020:43}. 
	Governed by quantum mechanical dipole selection rules ($\sigma^{\pm}$-transitions), OAM superposition states with uncommon symmetry properties have been generated \cite{Kerbstadt:2019:APX:1672583,Kerbstadt:2019:NC:658}. By combining photoionization using pulse-shaper generated polarization-tailored laser pulses with photoelectron tomography, unprecedented control of the 3D photoelecton angular distributions has been demonstrated.\\
	In this contribution, we demonstrate that a broad set of electron states can be generated by quantum interference using holographic electron diffraction in TEM and MPI. Experimental results for the preparation of OAM superposition states with both approaches are presented and analyzed in a unified theoretical description. Control of the phase structure of the OAM superposition states, as well as their symmetry and topological character are discussed.

	\section{Experiment}
	\label{Sec:experiment}
	
\begin{figure*}[t]
		\centering
		\includegraphics[width=\textwidth]{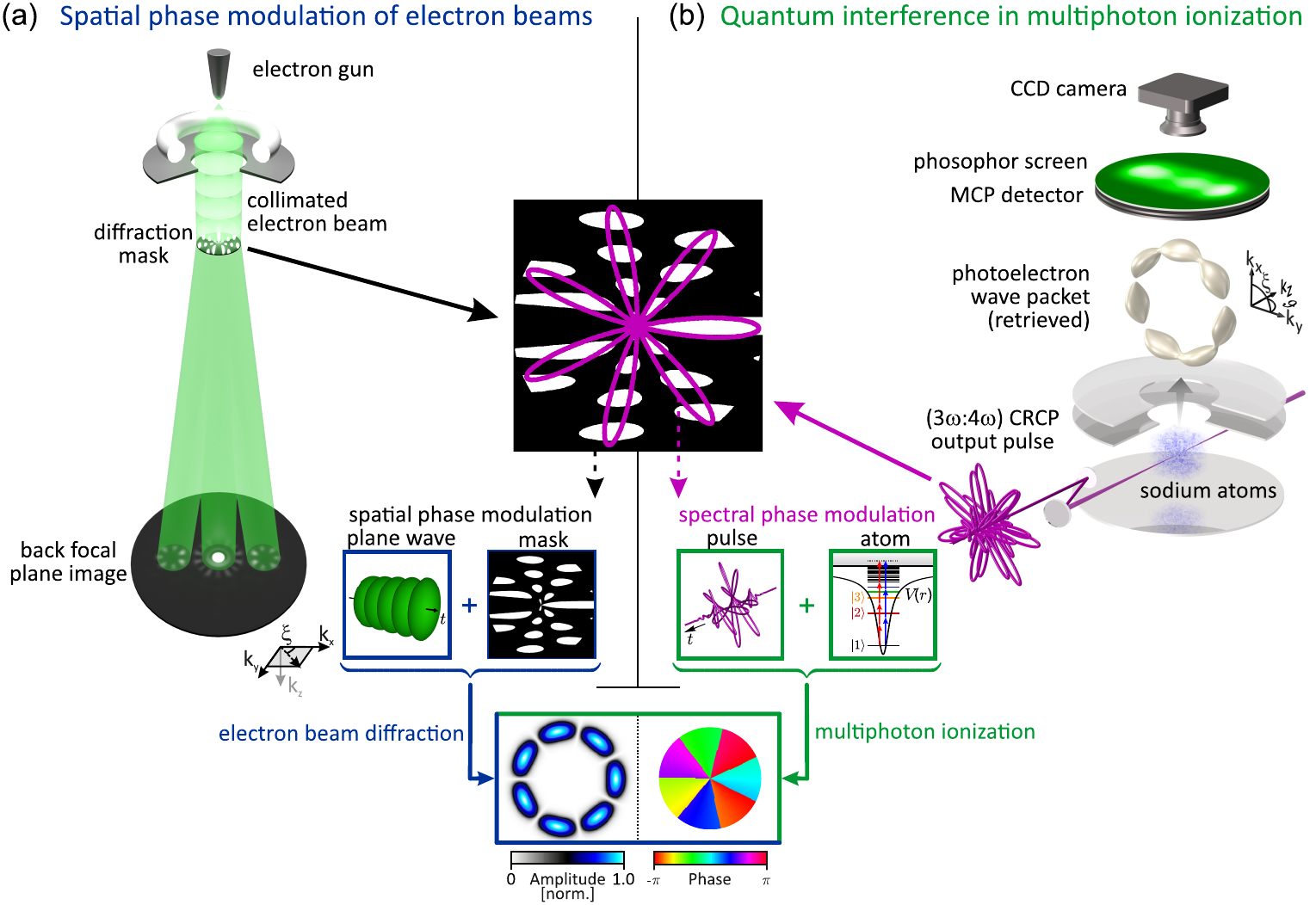}
		\caption{Schematic experimental setups. (a) In the SPM approach, a tailored holographic mask is illuminated by a spatially coherent electron beam in a transmission electron microscope. OAM superposition electron states are formed via far-field diffraction. (b) In the MPI approach, the interaction of a pulse-shaper generated ultrashort polarization-tailored supercontinuum with sodium atoms simultaneously drives two pre-selected ionization channels, leading to the generation of a superposition photoelectron state, which is detected in a VMI spectrometer. Both approaches result in electron states with a 7-fold rotationally symmetric electron density.}
		\label{fig:Scheme}
	\end{figure*}

	We investigate two complementary experimental approaches to generate and manipulate tailored OAM superposition electron states, utilizing spatial phase modulation (SPM) in a transmission electron microscope and multipath quantum interference from bichromatic MPI of atoms. A sketch of the experimental setups, highlighting the analogies of both approaches, is shown in figure~\ref{fig:Scheme}. The electron distributions of the generated states are detected in momentum-space, either in the far-field in the case of the SPM approach or by velocity map imaging (VMI) photoelectron spectroscopy.\\
	\noindent In general, OAM superposition states, also termed mixed-OAM states, are composed of two single OAM states with topological charges $\ell_1=-n$ and $\ell_2 =m$. In a two-dimensional space of polar momentum coordinates $(k,\xi)$, such OAM superposition states are generally described by:
	\begin{equation}\label{eq:superpos}
	\Psi(k, \xi) = G(k) \left(\beta e^{i m \xi} +  e^{-i n \xi} \right),	
	\end{equation}
	in which $G(k)$ is a real-valued radially depended function and $\beta=\beta_0 e^{i\gamma}$ a complex-valued superposition amplitude, with $\beta_0 \in \mathbb{R}_+$ controlling the relative amplitude of both components and $\gamma $ their relative phase. In the following, we consider $\beta_0 = 1$, a discussion of the azimuthal probability currents for superpositions with $\beta_0\neq 1$ is given in Sec.~\ref{Sec:results}. The superposition phase $\gamma$ is controlled by experimental means, either by the construction of the mask in SPM, or by the optical phases of the bichromatic field in MPI, as explained in Sec.~\ref{sec:techn_capabilities}.\\
	For a single OAM state $e^{im\xi}$ with integer $m \neq 0$, the electron density shows a dougnut-shaped, azimuthally symmetric distribution \cite{Cohen-Tannoudji:1991,Pengel:2017:PRL:053003}. The interference in the OAM superposition state described by Eq.\eqref{eq:superpos} leads to a reduced azimuthal symmetry. In contrast to the electron wave function $\Psi$, which belongs to the $C_1$ point group, the probability density of the superposition state is obtained as
	\begin{align} \label{eq:superpositionState}
	\vert \Psi(k,\xi) \vert^2 =  2 \vert  G(k) \vert^2 \left[ 1  + \cos\bigg([n+m] \xi + \gamma  \bigg) \right], 
	\end{align} 
	yielding an ($n+m$)-fold rotational symmetry \cite{Kerbstadt:2019:NC:658,Pengel:2017:PRL:053003}. 
	
	\subsection{Spatial phase modulation of electron beams}
	\label{sec:spatial_phase_mod_elec_beams}
	
\begin{figure*}[t]
		\centering
		\includegraphics[width=\textwidth]{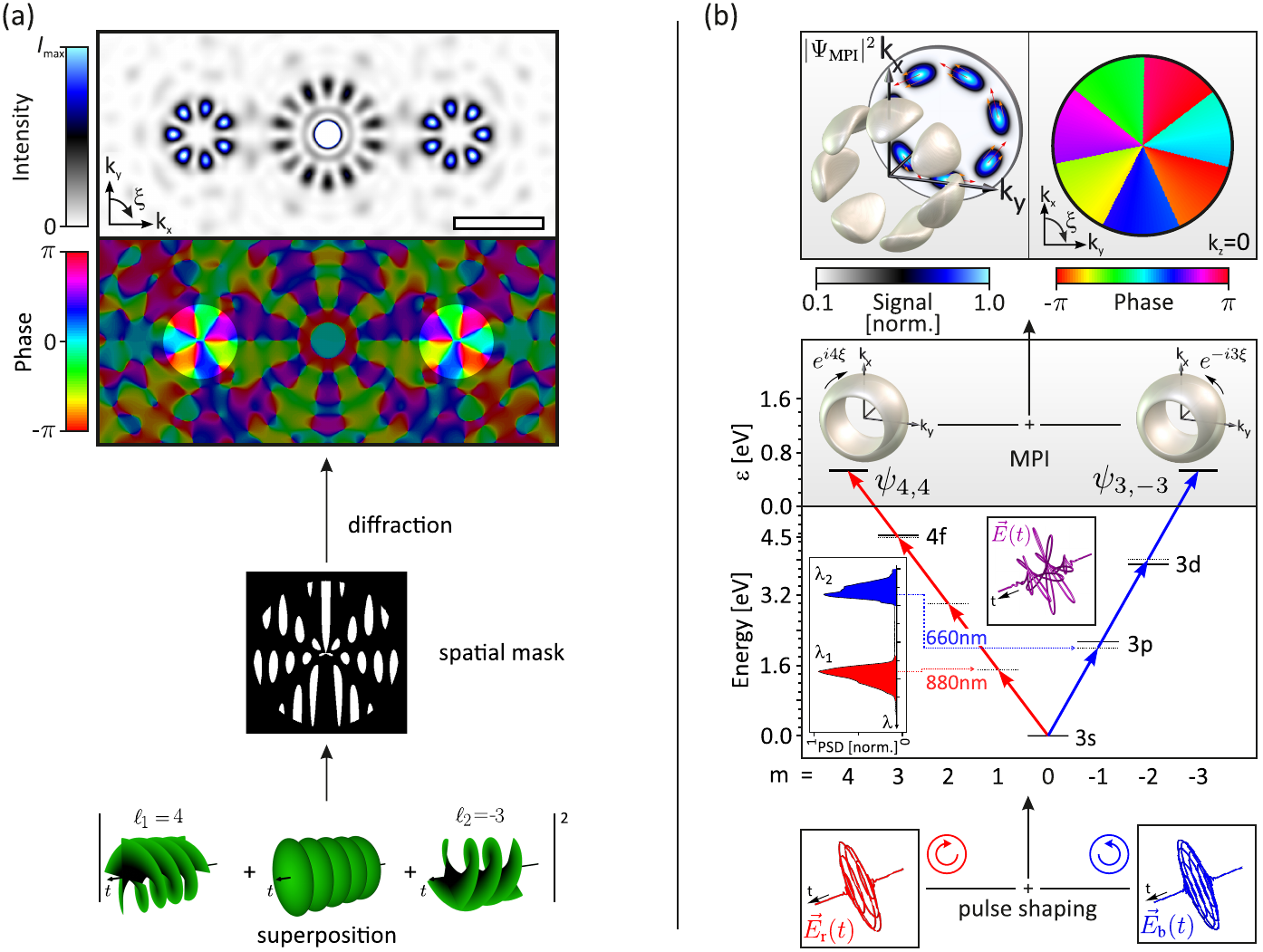}
		\caption{Concepts for the generation of OAM superposition electron states utilizing SPM and MPI. (a) In the SPM approach, a holographic mask (middle) is generated by mixing a plane-wave reference state with a target electron state, which is chosen here as a superposition of two OAM states (bottom). The calculated electron intensity and phase distribution in the far-field (top) reveal a mixed-OAM state with 7-fold rotational symmetry within the first diffraction orders (real-space mask diameter: $\SI{3.7}{\micro m}$, scale bar: $ \SI{10}{\micro m^{-1}}$). (b) A tailored CRCP bichromatic laser field (bottom) addresses predefined quantum pathways in the multiphoton excitation of sodium atoms (middle) and leads to a mixed-OAM photoelectron state composed of two torus-shaped single OAM states (top). The interference of the two quantum paths can be interpreted as a \emph{spectral double slit}, as opposed to the \emph{advanced spatial double slit} in (a).}
		\label{fig:Spatial_Modulation_Results}
	\end{figure*}

	For the generation of OAM electron beams in a TEM, we utilize a hologram-based approach \cite{Bazhenov.1990,Verbeeck:2010:Nature:301,Guzzinati.2013,Johnson.2020} (experimental sketch in figure~\ref{fig:Spatial_Modulation_Results}(a)), in which a diffraction mask $\mathcal{M}$ is derived from the superposition of an apertured plane wave $\text{circ}_R(r) e^{i k_0 x}$ and a real-space target electron wave function 
	\begin{align}
	\mathit{\Phi(r,\phi)}= \text{circ}_R(r) \cdot \left( e^{i m \phi} \cdot e^{i \kappa_\mathrm{SPM} }+ e^{-i n \phi} \right),
	\end{align}
	yielding the real-valued mask function
	\begin{align}\label{eq:mask_definition}
	\mathcal{M}(r,\phi) = \text{circ}_R(r) \cdot  \vert \left( e^{i m \phi} \cdot e^{i \kappa_\mathrm{SPM} }+ e^{-i n \phi} \right)  + e^{ik_0 x}\vert^2.
	\end{align}
	Here, $\text{circ}_R(r)$ is the circular aperture function with radius $R$ and $\kappa_\mathrm{SPM}$ a relative phase. By illuminating this mask with an electron plane wave, the mask's diffraction pattern $\Gamma(k_x,k_y)$ (cf. figure~\ref{fig:Scheme}(a)) is formed in the far-field:  
	\begin{align} \label{eq:FFT Maske}
	\Gamma(k,\xi) = \vert \mathcal{F}\lbrace\mathcal{M}  \rbrace\vert^2.
	\end{align}
	As shown in \ref{app:mask_calculation}, $\mathcal{F}\lbrace\mathcal{M} \rbrace$ consists of a central component and two side lobes at distances $\pm k_0$ in $k_x-$direction. Following the general principle of holography, the Fourier transform of the real-space target state, $\mathcal{F}\lbrace \vert \mathit{\Phi}(r,\phi)\vert^2 \rbrace$, and its complex conjugate are forming the side lobes. For a given reciprocal distance $k^{\mp}=k_\mathrm{eq}$ around the center of the first diffraction order of the reference wave, the phase behaviour of the Fourier-transformed target wave follows Eq.\eqref{eq:superpos}. Moreover, for $k^{\mp} \approx k_\mathrm{eq}$ the probability density of the superposition state can be approximated by
	\begin{equation}\label{eq:electron_density_SPM}
	\vert \Psi_{\text{SPM}}(k^{\mp},\xi^{\mp}) \vert^2 \approx 2 \vert G_{\text{SPM}}(k^{\mp}) \vert^2 \bigg[ 1 + \cos\big( \left[ n+m \right] \xi^{\mp} + \gamma   \big)  \bigg],
	\end{equation}
	in which $G_{\text{SPM}}(k^{\mp})$ is a function governed by the choice of the circular aperture radius, $k^{\mp}$ and $\xi^{\mp}$ denote shifted momentum coordinates and $\gamma= \kappa_\mathrm{SPM} \pm (m-n) \frac{\pi}{2}$ (cf. \ref{app:mask_calculation}).
	As an example, we show in figure~\ref{fig:Spatial_Modulation_Results}(a) the amplitude and phase of the numerical 2D-Fourier transform of a mask structure for $m=4$ and $n=3$.\\ 
	Experimentally, we utilized a binarized version of the calculated mask $\mathcal{M}$ (binarization threshold at half intensity maximum), which we fabricated by focused ion beam milling of a \SI{30}{nm} gold thin film on a silicon nitride membrane (\SI{15}{nm} thickness). In the cut-through sections of the mask, the electron wave is transmitted. Non-cut sections result in wave components scattered by large angles and are subsequently blocked by apertures. For the hologram, we chose $k_0 = \SI{15}{\micro m^{-1}}$ and an aperture diameter of $\SI{3.7}{\micro m}$. The mask is illuminated by an electron beam with large coherence length (coherence length: $d_{\text{c}}=\frac{\lambda}{2 \alpha}= \SI{1.74}{\micro m} $, angular spread: $2 \alpha= \SI{1.44}{\micro rad} $) formed in a transmission electron microscope (\emph{JEOL} JEM-2100F, $\SI{200}{kV}$ acceleration voltage). Using the post-specimen imaging lens system, the diffraction pattern of the mask is projected on the detector (\emph{GATAN} Orius SC600 charge-coupled device (CCD) camera; effective camera length: $\SI{300}{m} $), giving access to the spatial profile of the probability density of the electron state (cf. figure~\ref{fig:Scheme}).

	\subsection{Quantum interference in multiphoton ionization}
	\label{sec:spec_interf_elec_WP}
	
	In the ultrafast bichromatic MPI approach, sodium atoms are ionized with polarization-tailored bichromatic laser pulses \cite{Kerbstadt:2017:JMO:1010} to induce interference of pre-selected electron wave functions \cite{Kerbstadt:2017:NJP:103017,Bayer:2019:NJP:033001,Kerbstadt:2019:NC:658,Eickhoff:2020:PRA:013430}. Here, we consider atomic MPI (with $N_{\text{p}}$ photons) from the sodium 3s ground state (figure~\ref{fig:Spatial_Modulation_Results}(b)) using counter-rotating circularly polarized (CRCP) bichromatic propeller-type pulses (shown in figure~\ref{fig:Spatial_Modulation_Results} and \ref{fig:CoherentControl}) to generate OAM superposition states, yielding the photoemitted electron wave function in spherical momentum coordinates $(k,\xi,\vartheta)$ as
	\begin{align}\label{eq:Wave_function_MPI}
	\Psi_{\text{MPI}}(k,\xi,\vartheta) \propto \psi_{m,m}(k,\xi,\vartheta) \cdot e^{i \kappa_\mathrm{MPI}} + \psi_{n,-n}(k,\xi,\vartheta),
	\end{align}
	using 
	\begin{equation} \label{EQ8}
	\psi_{l, m}(k,\xi,\vartheta) = i^{N_{\text{p}}} \mathcal{R}_{l}(k) \mathcal{P}_{l ,m}[\cos(\vartheta)] e^{i m \xi},
	\end{equation}
	with the radial part $\mathcal{R}_{l}(k)$ of the continuum wave function (determined by the $N_{\text{p}}$-th order spectrum of the laser pulse), the associated Legendre polynomials $\mathcal{P}_{l ,m}[\cos(\vartheta)]$ and the phases $i^{N_{\text{p}}}$ from $N_{\text{p}}$-th order perturbation theory (for details see \ref{app:MPI} and \cite{Kerbstadt:2019:NC:658}). The relative phase $\kappa_\mathrm{MPI}$ is adjusted by optical phases introduced by the pulse shaper \cite{Weiner:2000:RSI:1929,Kerbstadt:2017:JMO:1010} and the polarization-rotation optics. The two wavefunction components in Eq.\eqref{eq:Wave_function_MPI} originate from the 3- and 4-photon-ionization channels and interfere in the same energy window (about \SI{0.5}{eV}) of the continuum states, as depicted in figure~\ref{fig:Spatial_Modulation_Results}(b), resulting in the superposition of two OAM electron states $\psi_{n,-n}(k,\xi,\vartheta)$ and $\psi_{m,m}(k,\xi,\vartheta)$ with topological charges $\ell_1 = -n$ and $\ell_2 =m$, respectively. The target states are addressed via $\sigma^{\pm}$-transitions due to the left and right circular polarized (LCP/RCP) laser electric fields, for which the respective quantum numbers are determined by selection rules (${\Delta l = 1}$, ${\Delta m = \pm 1}$). As discussed in \ref{app:MPI}, around $\vartheta= \frac{\pi}{2}$, the momentum-space wave function in Eq.\eqref{eq:Wave_function_MPI} factorizes and follows Eq.\eqref{eq:superpos}. Hence, the electron wave function exhibits a phase structure in the ($k_x,k_y$)-plane similar to the spatially phase modulated electron beam in the TEM. As a consequence, the resulting electron density around $\vartheta = \frac{\pi}{2}$ can be written as
	\begin{equation}\label{eq:electron_density_MPI}
	\vert \Psi_{\text{MPI}}(k,\xi,\vartheta) \vert^2 \approx 2 \vert G_{\text{MPI}}(k) \vert^2 \bigg[ 1 + \cos\big( \left[ n+m \right] \xi + \gamma   \big)  \bigg],
	\end{equation}
	in which $G_{\text{MPI}}(k) = \mathcal{R}_{n}(k) \mathcal{P}_{n ,-n}[\cos(\vartheta)]$ and $\gamma = \kappa_\mathrm{MPI} + (m-n)\frac{\pi}{2} + m \pi $ in full analogy to Eq.\eqref{eq:electron_density_SPM}.\\
	For the experimental implementation of MPI-based electron state generation and characterization, we combine bichromatic polarization pulse shaping (more details of the experimental setup are given in \cite{Kerbstadt:2017:OE:12518}) with a VMI based photoelectron tomography \cite{Eppink:1997:RSI:3477,Wollenhaupt:2009:APB:647,Wollenhaupt:2013:CPC:1341} sketched in figure~\ref{fig:Scheme}. Near-infrared femtosecond pulses from a multipass chirped pulse amplifier (\emph{FEMTOLASERS} Rainbow 500, CEP4 module, Femtopower HR, $\SI{3}{kHz}$ repetition rate; $\lambda_0=\SI{790}{nm}$, \SI{1.0}{mJ} pulse energy) with actively stabilized carrier-envelope phase (CEP) are employed to seed a neon-filled hollow-core fiber for the generation of an octave-spanning white light supercontinuum (WLS). The white light pulses are modulated in the frequency domain using a home-built $4f$ polarization pulse shaping setup \cite{Brixner:2001:OL:557,Weiner:2011:OC:3669,Kerbstadt:2017:JMO:1010}, which consists of a dual-layer liquid crystal spatial light modulator (LC-SLM; \emph{Jenoptik} SLM-640d) in combination with a custom polarizer. For the conversion from linear to counter-rotating circular polarization, we utilize a superachromatic $\lambda/4$-waveplate at the shaper output. The generated bichromatic \bichrom{3}{4} CRCP field (cf. inset figure~\ref{fig:Spatial_Modulation_Results}(b)), consisting of a \emph{red} ($\lambda_{\text{r}}=\SI{880}{nm}$, $\Delta t_{\text{r}} \approx \SI{25}{fs}$, LCP) and \emph{blue} ($\lambda_{\text{b}}=\SI{660}{nm}$, $\Delta t_{\text{b}} \approx \SI{25}{fs}$, RCP) component, is CEP-stabilized using an external active stabilization loop \cite{Kerbstadt:2018:PRA:063402}. To this end, an f-2f interferometer is implemented, fed by an additional \bichrom{}{2}-field extracted from the spectral edges of the WLS. The bichromatic fields are focused into the interaction region of a VMI spectrometer (peak intensity $I_0\approx \SI{2e12}{W/cm^2}$) filled with sodium vapor. The photoelectron wave packets created by atomic MPI are projected onto a position-sensitive 2D detector consisting of a Chevron micro-channel-plate (MCP) and a phosphor screen and are recorded by a CCD camera. For tomographic reconstruction of the 3D photoelectron momentum distribution (PMD) the input pulse sequence is rotated around the laser propagation direction by using a superachromatic \mbox{$\lambda/2$-waveplate} \cite{Wollenhaupt:2009:APB:647,Wollenhaupt:2013:CPC:1341}. Each PMD was retrieved from 45 projections, measured with an angular step size of $\delta\phi=4^{\circ}$, employing the Fourier slice algorithm \cite{Kak:1988:1}. Note that the VMI detection scheme, results in a projection of the PMD which scales linearly in the radial direction with the electron energy. Therefore, the displayed MPI results show a non-linear radial dependence, as compared to the SPM results.

	\section{Results and discussion}
	\label{Sec:results}
	
	\subsection{Generation of mixed-OAM states}
	
\begin{figure*}[t]
		\centering
		\includegraphics[width=\textwidth]{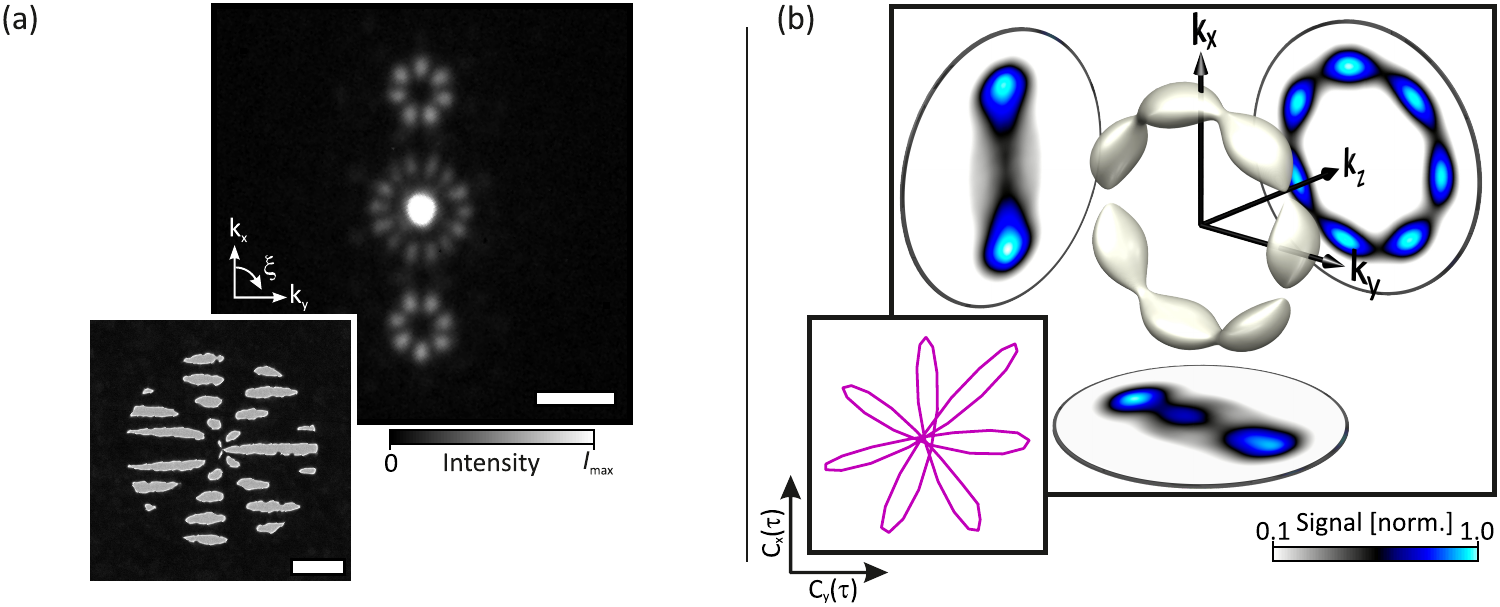}
		\caption{Experimental mixed-OAM electron states. (a) The electron intensity distribution formed by far-field diffraction of a holographic spatial gold mask reveals a 7-fold rotationally symmetric electron density in the diffraction sidebands ($\SI{3.7}{\micro m}$ aperture diameter, $k_0= \SI{15}{\micro m^{-1}}$, ($m=4$, $n=3$), scale bar = $ \SI{10}{\micro m^{-1}} $). (Inset) Transmission electron micrograph of the employed holographic mask ($\SI{30}{nm}$ gold on $\SI{15}{nm}$ silicon nitride membrane; scale bar: $\SI{1}{\micro m} $). (b) The measured electron density projections and the corresponding 3D tomographic reconstruction of mixed-OAM electron states, generated from MPI of sodium atoms with tailored \bichrom{3}{4} laser fields (${\varphi_{\text{r}} = \varphi_{\text{b}} = \varphi_{\text{ce}} =0}$), reveals also a 7-fold rotationally symmetric photoelectron distribution. (Inset) The experimentally determined polarization profile of the bichromatic laser field.}
		\label{fig:CoherentControl}
	\end{figure*}

	The experimental OAM superposition states generated by SPM and MPI are compared in figure~\ref{fig:CoherentControl} for ${m=4}$ and ${n=3}$. Both, the intensity distribution in the first-order sidebands in SPM (figure~\ref{fig:CoherentControl}(a)) and the MPI results (figure~\ref{fig:CoherentControl}(b)), show a 7-fold rotationally symmetric flower-petal-like structure, as expected from Eq.\eqref{eq:superpositionState}. Due to the phase singularity of OAM states at ${k^{\pm}=0}$, the intensity of the OAM superposition vanishes at the center in both cases. We note that the radial behaviors of the SPM- and MPI-generated states, depending on the mask aperture via $G_{\text{SPM}}(k)$ and the radial part of the atomic wave function in $G_{\text{MPI}}(k)$ respectively, are different in the SPM and MPI cases. Furthermore, in the TEM approach the generated mixed-OAM states exhibit no $k_z$-dependence, whereas in MPI an additional $\vartheta$-dependence is introduced via $\mathcal{P}_{n ,-n}[\cos(\vartheta)]$.
	It is instructive to further consider the symmetries of the spatial mask $\mathcal{M}$ and the electric field distribution in the MPI light field. The mask $\mathcal{M}$ can be explicitly expressed as
	\begin{align}\label{eq:Mask_explicit}
	&\mathcal{M} = \text{circ}_R(r) \bigg[ 3 + 2 \cos([n+m] \phi + \kappa_\mathrm{SPM})  \notag \\
	&\phantom{\mathcal{M} = \text{circ}_R(r) \bigg[ 3} +4\cos\left(\frac{(n-m)\phi-\kappa_\mathrm{SPM}+2k_0x}{2} \right)  \cos\left(\frac{(m+n)\phi+\kappa_\mathrm{SPM}}{2} \right) \bigg] 
	\end{align}
	using Eq.\eqref{eq:mask_definition}. Despite the low symmetry of the mask function, the overall lobular structure of the mask, depicted in figure~\ref{fig:CoherentControl}(a), follows a 7-fold rotational symmetry due to the second term in Eq.\eqref{eq:Mask_explicit}. The high-frequency mask components (third term in Eq.\eqref{eq:Mask_explicit}) are formed by a spatial beating between the target wave function and the reference plane wave, yielding the OAM sidebands in the diffraction pattern of the mask. In the MPI scenario, the time-dependent CRCP optical electric field $\vec{E}(t)$ with commensurable frequencies $n\omega$ and $m\omega$ and equal envelopes $\mathcal{E}_0(t)$ is given by
	\begin{equation}\label{eq:el_field}
	\vec{E}(t) = \sqrt{2}\mathcal{E}_0(t)   \cos \left( \frac{(n+m)}{2} \omega t \right)  \begin{pmatrix}
	\cos \left( \frac{(n-m)}{2} \omega t \right)\\ 
	\sin \left( \frac{(n-m)}{2} \omega t \right)
	\end{pmatrix},	
	\end{equation} 
	and contains an expression describing the beating, similar to the corresponding expression in Eq.\eqref{eq:Mask_explicit}. Projecting the trace of the temporally evolving electric field vector onto the transverse polarization plane, results in a propeller-shaped curve with $\mathcal{S}_{\text{opt}}$-fold rotational symmetry, given by $	\mathcal{S}_{\text{opt}} = (n+m)/\gcd(n,m)$,  
	where $\gcd(n,m)$ denotes the greatest common divisor of $n$ and $m$ \cite{Kerbstadt:2019:NC:658}. A measured polarization profile of the generated phase-stable \bichrom{3}{4} CRCP field is depicted in the inset of figure~\ref{fig:CoherentControl}(b) (for more experimental details see \cite{Kerbstadt:2017:OE:12518}), highlighting the similarity to the corresponding mask structure shown in the inset of figure~\ref{fig:CoherentControl}(a). 
	The close similarity suggests to interpret the diffraction mask in figure~\ref{fig:Spatial_Modulation_Results}(a) as an \emph{advanced double slit} analogously to the \emph{spectral double slit} \cite{Shapiro:2003} in the MPI framework (cf. figure~\ref{fig:Spatial_Modulation_Results}(b)).\\
	For a more detailed discussion of the topological properties of the OAM superposition state from Eq.\eqref{eq:superpos}, we consider the probability current $\vec{j}$ of the evolved real-space electron wave function \cite{Winter:2006:OC:285}, yielding \cite{Cohen-Tannoudji:1991,Bliokh:2017:PR:1,Kerbstadt:2019:APX:1672583,Kerbstadt:2020:43,Bayer:2020:PRA:013104}
	\begin{align}\label{eq:psicurrent_N1_N2}
	\vec{j} &= \frac{\hbar}{m} \Im 	\left[\Psi^* \vec{\nabla} \Psi\right] \propto  \vert \Psi \vert^2 \frac{m - n}{2} \vec{e}_{\xi} + \mathcal{O}(\beta_0 -1)  \approx \rho \frac{m-n}{2} \vec{e}_{\xi},
	\end{align}
	at a certain radius $k$, with the probability density $\rho$ and $\mathcal{O}(\beta_0 -1)$ denoting terms proportional to ${(\beta_0 -1)}$. Although the resulting electron density is a static structure, i.e. a standing wave, $\vec{j}$ does not vanish for $n\neq m$. The probability current shown in the inset of figure~\ref{fig:Spatial_Modulation_Results} is curling around the center of the structure in azimuthal direction $\vec{e}_{\xi}$ with an angular spatial frequency $\omega_{\xi} \propto \frac{m-n}{2}$ and an amplitude determined by $\rho$.\\
	For MPI, this observation is rationalized by the fact, that during photoionization, the probability current is driven by the electric field. The angular frequency of the laser electric field of a propeller-pulse with equal field amplitudes is given by $\omega_{\Phi} = \frac{m-n}{2} \omega$ which does not change within the pulse \cite{Reich:2016:PRL:133902,Mauger:2016:JPB:10LT01} (cf. \ref{app:electric_field}) and therefore $\omega_{\Phi}(m,n) \propto \omega_{\xi}(m,n)$. In the single color case both angular frequencies vanish, since ${m=n}$ \cite{Pengel:2017:PRA:043426,Pengel:2017:PRL:053003,Kerbstadt:2019:NC:658}. 
	In general, the topological charge of the superposition state, $\ell(n,m;\beta_0)$ \cite{Bliokh:2017:PR:1} is a function discontinuously depending on the parameter $\beta_0$ (cf. \ref{app:topo_charge}). In the specific case of $\beta_0=1$, the topological charge takes a fractional value $\ell = \frac{m - n}{2}$, leading to a discontinuous topological charge in the experiment, depending on the respective amplitudes of the different OAM states in the superposition. For $\beta_0 \neq 1$, the topological charge of the superposition state is given by the topological charge of the OAM state with the larger weight in Eq.\eqref{eq:superpos}. However, small variations around $\beta_0 =1$ do not affect the wave packet structure since $\Psi$ and $\vec{j}$ are continuous functions of $\beta_0$.\\
	Finally, we note that for single OAM states $e^{im\xi}$ the topological charge $m$ can be experimentally determined by applying an astigmatic defocus and counting the resulting number of intensity minima \cite{Kotlyar:2017:AO:4095,Shutova:2017:PLA:408}. A similar approach for the OAM superposition states only leads to a complex interference pattern with the number of intensity minima not directly connected to the topological charge.
	
	\begin{figure*}[t]
		\centering
		\includegraphics[width=\textwidth]{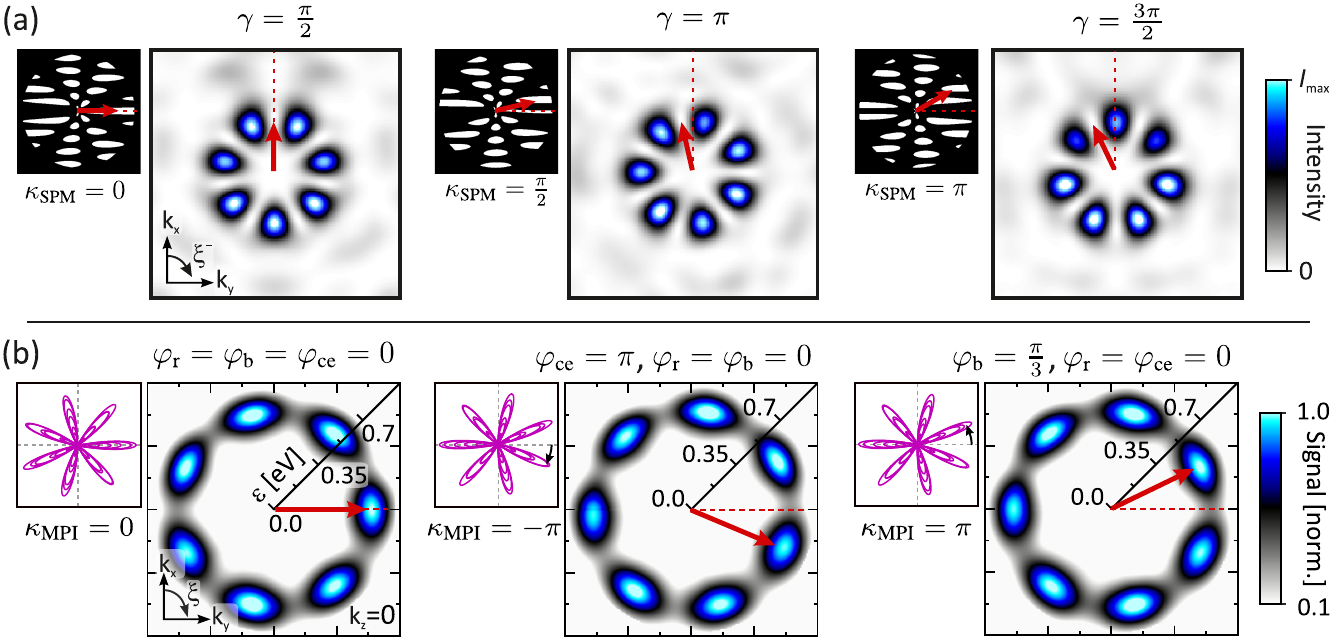}
		\caption{Phase control of OAM superposition electron states. (a) Calculated diffraction masks ($\SI{3.7}{\micro m}$ aperture diameter, $k_0= \SI{15}{\micro m^{-1}}$) and corresponding far-field electron intensities for different azimuthal phases~$\kappa_\mathrm{SPM}=0,\, \pi/2, \,\pi$. The magnified first diffraction order (bottom), shows a rotation of $\Gamma^{(-1)}$ by $\kappa_\mathrm{SPM}/7$. Field-of-view in zoom-in: $\SI{15}{\micro m^{-1}}$. (b) Measured photoelectron density in the MPI approach for different optical phases $\varphi_{\text{r}} = \varphi_{\text{b}} = \varphi_{\text{ce}}=0$ (left), $\varphi_{\text{r}} = \varphi_{\text{b}} =0$ and $\varphi_{\text{ce}}= \pi$ (middle) as well as $\varphi_{\text{r}} = \varphi_{\text{ce}} =0$ and $\varphi_{\text{b}} = \frac{\pi}{3}$ (right)}.
		\label{fig:Simulation_Rotation}
	\end{figure*}

	\subsection{Manipulation and control of OAM superposition states}
	\label{sec:techn_capabilities}
	
\begin{figure*}[t]
		\centering
		\includegraphics[width=\linewidth]{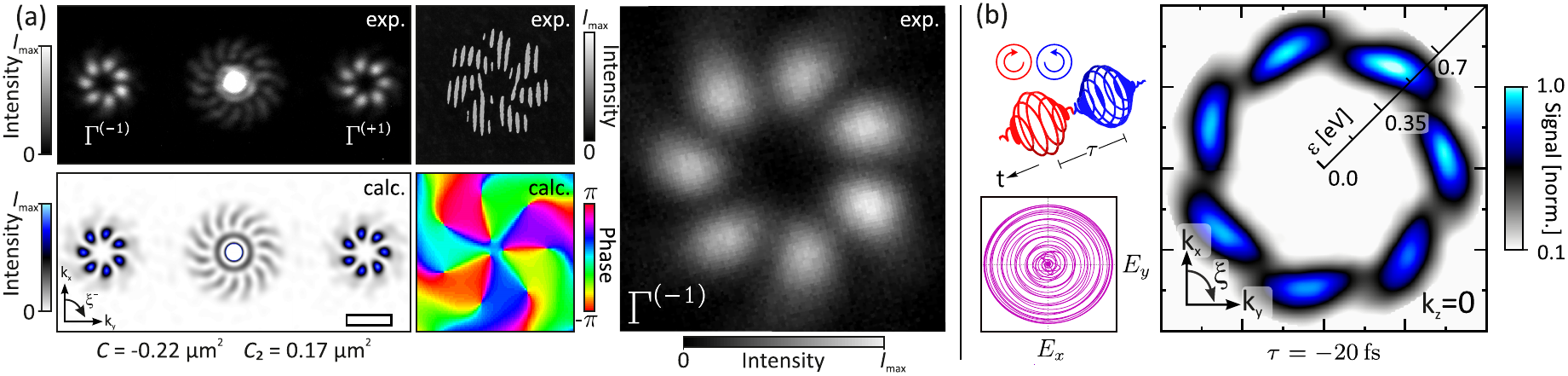}
		\caption{Generalized mixed-OAM electron states. OAM superposition states are accessible both in the SPM and MPI approach utilizing adapted mask geometries or time-delayed bichromatic ionization pulses, respectively. (a) Experimentally measured (top left and zoom in for $\Gamma^{(-1)}$ (right)) and calculated (bottom left) far-field electron intensity scattered from an adapted mask. For the mask manufacturing (TEM image (top center): mask diameter: $\SI{5.3}{\micro m}$), a phase-factor $e^{iCk^2}$ and a Gaussian envelope $e^{-\frac{1}{2}C_2 k^2}$ were incorporated ($C= \SI{-0.22}{\micro m^2}$, $C_2= \SI{0.17}{\micro m^2}$, see text and \ref{app:mask_calculation} for details). (bottom center) The calculated phase distribution exhibits a spiral-shaped radial structure. (b) The measured photoelectron momentum distribution of sodium atoms using a \bichrom{3}{4} pulse sequence with a time-delay $\tau = \SI{-20}{fs}$ applied to the \emph{blue} pulse, exhibits tilted lobes in the flower-petal structure. The temporally delayed pulse sequence and the $E_x$-$E_y$ projection of the corresponding laser electric field's polarization profile are shown in the inset. Both for the SPM and MPI approach, the $k$-dependent phase factor in the OAM superposition states results in the lobes of the electron density profile being inclined relative to the radial direction.}
		\label{fig:Radial_Phase}
	\end{figure*}

	The orientation of the flower-petal-like electron density is determined by the phase $\gamma$ (cf. Eq.\eqref{eq:superpositionState}). In the experimental SPM approach, $\gamma$ is set by the factor $\kappa_\mathrm{SPM}$ in the mask design. In figure~\ref{fig:Simulation_Rotation}(a), we show the simulated electron density profile in the diffraction pattern for $\kappa_\mathrm{SPM}=0,\, \pi/2$ and $\pi$, corresponding to $\gamma= \pi/2,\, \pi$ and $3\pi/2$. A rotation of the first-order sideband~$\Gamma^{(-1)}$ around the respective center by an angle $\gamma/(n+m)$ is visible (cf. Eq.\eqref{eq:electron_density_SPM}). According to Friedel's law applicable for real-valued masks, the diffraction pattern remains inversion symmetric independent of $\kappa_\mathrm{SPM}$ \cite{Friedel.1913,Juchtmans.2016} (cf. figure~\ref{fig:Radial_Phase}(a)). Notably, the approach does not correspond to a simple mask rotation, which would cause the whole diffraction pattern to rotate around its center at $k=0$.\\ 
	Similarly, in the MPI approach, $\gamma$ is controlled by optical phases via the parameter $\kappa_\mathrm{MPI}$ (cf. Eq.\eqref{eq:electron_density_MPI}), given by $\kappa_\mathrm{MPI} = -m \varphi_{\text{r}} +n \varphi_{\text{b}} - (m-n) \varphi_{\text{ce}} + (m+n)\zeta$ (for details see \ref{app:electric_field}), with the respective relative phases $\varphi_{\text{r/b}}$ of the \emph{red} and \emph{blue} component of the bichromatic laser field, the carrier-envelope phase $\varphi_{\text{ce}}$ and the relative angle $\zeta/2$ of the $\lambda/2$-waveplate \cite{Kerbstadt:2019:NC:658}. Similar to the SPM approach, the resulting photoelectron density is rotated by an angle of $\kappa_\mathrm{MPI}/(n+m)$. To highlight the analogy, three experimental examples of rotational phase control are depicted in figure~\ref{fig:Simulation_Rotation}(b) for the optical phases ${\varphi_{\text{r}} = \varphi_{\text{b}} = \varphi_{\text{ce}}=0}$ (left), ${\varphi_{\text{ce}}= \pi}$ (middle) and ${\varphi_{\text{b}}= \frac{\pi}{3}}$ (right). Furthermore, the control of the spatial rotation is not only accessible via relative spectral phases of the field but also by the adjustment of a $\lambda/2$-waveplate resulting in a rotation of the whole bichromatic laser field in the polarization plane (see e.g. Eq.(6) in \cite{Wollenhaupt:2009:APB:647}). In fact, this feature is crucial for the tomographical reconstruction of the full 3D PMD \cite{Bayer:2019:NJP:033001,Kerbstadt:2019:APX:1672583,Smeenk:2009:JPB:185402,Itatani:2004:Nature:867}. Importantly, not only optical but also quantum phases accumulated during the photoemission process result in a rotation of the detected electron density \cite{Eickhoff:2020:PRA:013430,Bayer:2020:PRA:013104}. Hence, the MPI approach further enables to study imprinted time-dependent dynamics of quantum systems, e.g., due to spin-orbit coupling \cite{Bayer:2019:NJP:033001} or Rydberg states \cite{Kerbstadt:2019:PRA:013406,Eickhoff:2020:PRA:013430}. Moreover, it has been demonstrated that the photoelectron wave packet's symmetry is also controlled via the laser intensity as the interaction evolves from the perturbative to the strong-field regime \cite{Li:2018:OE:878,Pengel:2017:PRA:043426}.\\
	Recently, spiral shaped electron wave packets (electron vortices) in MPI \cite{Pengel:2017:PRA:043426,Pengel:2017:PRL:053003,Kerbstadt:2019:NC:658,Bayer:2020:PRA:013104,Kerbstadt:2019:APX:1672583,Kerbstadt:2020:43} have attracted significant attention. These vortices have a $k$-dependent phase due to the time-evolution of the wave packet \cite{Wollenhaupt:2002:PRL:173001}.\\
	In electron wave optics, spatial control of electron beams by magnetic lenses is conveniently described by spatial phase masks applied in the back focal plane of the imaging lens \cite{Erni.2015,Zuo.2017}. The corresponding phase masks typically exhibit $k$-dependent phase functions. For example, the effect of a non-aberrated circular symmetric magnetic lens can be described by a phase mask $e^{i C k^2}$ with ${C \in \mathbb{R}}$, resulting in a converging parabolic wavefront \cite{Erni.2015,Zuo.2017}. 
	In the holographic TEM approach utilized here, an equivalent $k$-dependent phase function $\gamma(k)$ can be imprinted onto the diffracted electron wave by choosing a more sophisticated mask design. In particular, the holographic mask is calculated by applying the (inverse) Hankel transform of the respective targeted $k$-dependence in position space. For this reason we choose the $m$-th order normalized Hankel transform of $ e^{iC k^2} e^{-\frac{1}{2}C_2 k^2}$ and the $n$-th order normalized Hankel transform of $e^{-\frac{1}{2}C_2 k^2}$ as complex amplitudes for the respective partial states $e^{i m \phi}$ and $e^{- i n \phi}$, respectively (for more details see \ref{app:mask_calculation}). Experimental and calculated results for the far-field electron diffraction pattern and phase-distribution are shown in figure~\ref{fig:Radial_Phase}(a), indeed exhibiting a spiral-shaped electron distribution. \\
	In the MPI approach, a similar $k$-dependent phase appears during the time-evolution of the wave packets when an additional time-delay $\tau$ (applied to the \textit{blue} pulse) is introduced between the two spectral components in the bichromatic field \cite{Pengel:2017:PRL:053003,Pengel:2017:PRA:043426,Kerbstadt:2019:NC:658}. This delay yields $\gamma(k)=\gamma_0 + \frac{\hbar \tau}{2 m} k^2$, where $\gamma$ of Eq.\eqref{eq:superpos} is denoted as $\gamma_0$.
	The resulting photoelectron density is displayed in figure~\ref{fig:Radial_Phase}(b), showing a pronounced tilt of the petal lobes relative to the radial direction. The time-delay leads to a $\tau$-dependent radial component of the probability current $\vec{j}$ along with a radial dependence of the interference term, giving rise to the spiral shape \cite{NgokoDjiokap:2015:PRL:113004}. The tilt angle of the lobes increases, for larger time delays. In contrast to the PMD, for sufficiently large time-delays, i.e. when the delay exceeds the pulse duration, the spectral components in the bichromatic fields are temporally separated, such that the polarization profile is circular and does not show the propeller-type structure in the polarization plane. Hence, in the multiphoton regime the symmetry of the wave packet is completely described by the quantum interference of states with different angular momenta and not fully determined by the optical field structure. For a vanishing time-delay the photoelectron wave packet symmetry generally maps the field symmetry when the difference between both photonicities ($n$ and $m$) equals $1$, i.e., when the wave packet exhibits an odd ($n$+$m$)-fold rotational symmetry, as discussed in Ref. \cite{Kerbstadt:2019:NC:658,Kerbstadt:2020:43,Kerbstadt:2019:APX:1672583}. While the MPI technique enables radial control via the $N_{\text{p}}$-th order laser electric field's spectrum, the holographic TEM approach can be extended to other radial phase dependencies, allowing advanced transverse control of the resulting electron distribution. \\

	\section{Conclusion and Outlook} 
	\label{sec:conclusion}
	
	In this paper, we presented the generation and manipulation of orbital angular momentum (OAM) superposition electron states using tailored holographic spatial masks in a transmission electron microscope (TEM) and shaper-generated bichromatic laser pulses for atomic multiphoton ionization (MPI). Both approaches, were interpreted in the physical picture of an \emph{advanced double-slit} in either the spatial or spectral domain, resulting in electron density distributions with a 7-fold rotational symmetry. Further control aspects, including a rotational or radial phase control, were demonstrated, unifying the theoretical concepts commonly employed in electron wave optics and in the ultrafast coherent control of photoelectron wave packets, respectively. So far, we focussed on a comparison of both approaches. However, fascinating perspectives arise when both experimental techniques are combined. For example multiphoton photoemission from atomic systems may also be useful for the generation of phase structured electron wavefronts in transmission electron microscopy, similar to recent work in ultrafast electron diffraction \cite{McCulloch:2011:NP:785,McCulloch.2013,Engelen:2013:NC:1}. In addition, the MPI technique can be extented to molecular MPI \cite{Yuan:2016:PRA:053425,Yuan:2017:JPB:124004,Djiokap:2018:PRA:063407,Chen:2020:PRA:033406} giving rise to molecular OAM states in photoionization. Since, the rotation and the symmetry of the generated OAM states is affected by the atomic or molecular system itself, such an approach enables the precise measurement of quantum phase shifts, similar to \cite{Eickhoff:2020:PRA:013430}, and Stark-shifts \cite{Li:2018:OE:878}. Finally, the uncommon 7-fold rotationally symmetric electron density of the OAM states demonstrated here may be useful for attaining enhanced sensitivity in scanning low-loss electron energy spectroscopy of plasmonic particles with related symmetries.

	\ack
	Financial support by the Deutsche Forschungsgemeinschaft via the priority programme SPP1840 QUTIF and by the Volkswagen Foundation as part of the Lichtenberg Professorship ``Ultrafast nanoscale dynamics probed by time-resolved electron imaging'' is gratefully acknowledged. The electron microscopy service unit of the University of Oldenburg is acknowledged for their technical support in TEM operation.
	
	\section*{Data availability}
	The data that support the findings of this study are available upon request from the authors.

	\appendix
	\section{Diffraction pattern of holographic TEM mask}
	\label{app:mask_calculation}
	
	In this section, we derive the far-field diffraction of the tailored TEM mask, containing OAM superposition states. Following the holographic mask approach in \cite{Verbeeck:2010:Nature:301}, we construct a mask function given by 
	\begin{equation}\label{eq:Append_Mask_design}
	\mathcal{M}(r,\phi) = \text{circ}_R(r) \vert  e^{i m \phi}e^{i \kappa_\mathrm{SPM}} +  e^{-i n \phi} + e^{i k_0 x} \vert^2,
	\end{equation}
	with variables as introduced in the main text. Equation~\eqref{eq:Append_Mask_design} can be written as a sum of four terms, $\mathcal{M} = \mathcal{M}_0 + \mathcal{M}_1 + \mathcal{M}_2 + \mathcal{M}_3$, with
	\begin{align}\label{eq:mask_three_terms}
	&\mathcal{M}_0 = 3~ \text{circ}_R(r), \\
	&\mathcal{M}_1 = \text{circ}_R(r) \left( e^{i(m+n)\phi}e^{i \kappa_\mathrm{SPM}} + e^{-i(m+n)\phi}e^{-i \kappa_\mathrm{SPM}} \right),\\
	&\mathcal{M}_{2,3} = \text{circ}_R(r) e^{\pm i k _0 x} \left( e^{\pm in\phi} + e^{\mp im\phi}e^{\mp i \kappa_\mathrm{SPM}} \right). \label{eq:mask_three_terms2}
	\end{align}
	Rearranging~$\mathcal{M}$ in terms of cosine functions yields Eq.\eqref{eq:Mask_explicit}.
	%
	The far-field diffraction of the electron wave transmitted through the mask is described by the mask's 2D Fourier transform 
	\begin{equation}
	\mathcal{F}\lbrace \mathcal{M} \rbrace(k_x,k_y) = \int_{\mathbb{R}^2} \mathcal{M}(x,y)~ e^{-i (k_x x + k_y y)} \d x \d y,
	\end{equation}
	which is given in polar coordinates $(r,\phi)$ by \cite{Baddour:2009:JOSAA:1767}
	\begin{equation}
	\mathcal{F}\lbrace f \rbrace (k , \xi) = \int_{0}^{\infty} \int_{- \pi}^{\pi} f(\vec{r}) ~e^{-i r k \cos(\xi - \phi )} r \d r \d \phi,
	\end{equation}
	with $k = \vert \vec{k} \vert$ and the polar angle of the transverse momentum~$\xi = \arctan(k_y/k_x)$. For non-radial symmetric functions $f(\vec{r})$, one can use an angular Fourier decomposition \cite{Baddour:2009:JOSAA:1767}
	\begin{equation}\label{eq:decomposition_fourier}
	f(\vec{r}) = f(r,\phi) = \sum_{q =-\infty}^{\infty} f_q(r) e^{i q \phi}.
	\end{equation}
	With the 2D Fourier transform of Eq.\eqref{eq:decomposition_fourier} 
	\begin{equation}
	\tilde{f}(\vec{k}) = \sum_{q=- \infty}^{\infty} 2 \pi i^{-q} e^{i q \xi} \int_{0}^{\infty} f_q(r) J_q (k r) r \d r,
	\end{equation}
	we directly obtain the Fourier transformed components of the diffraction mask in Eqs.\eqref{eq:mask_three_terms}-\eqref{eq:mask_three_terms2}
	\begin{align} \label{M0}
	&\tilde{\mathcal{M}_0} = 6\pi \mathcal{I}_0(k) = 6 \pi \frac{R}{k} J_1(k R), \\
	&\tilde{\mathcal{M}_1} = 4 \pi \mathcal{I}_{(m+n)}(k)(-i)^{(m+n)} \cos \big( (m+n)\xi + \kappa_\mathrm{SPM} \big),   \\
	&\tilde{\mathcal{M}}_{2,3} = 2 \pi \big( i^{-m} e^{\mp i m \xi^\mp}e^{\mp i \kappa_\mathrm{SPM}} \mathcal{I}_m(k^\mp) + i^{-n}  e^{\pm in \xi^\mp} \mathcal{I}_n(k^\mp)  \big). \label{M3}
	\end{align}
	We note that $\tilde{\mathcal{M}}_{2,3}$ accounts for both side lobes in the diffraction pattern, which are equal to the Fourier transform (or its complex conjugate) of the real-space target function, shifted by $-k_0$ ($+k_0$) in the $k_x$-direction in reciprocal space, respectively. In deriving Eqs.\eqref{M0}-\eqref{M3}, we used the identity $J_{-n}(k r) = (-1)^n J_n(k r)$ and a short hand notation for the $n$-th order Hankel transform of $\text{circ}_R(r)$
	\begin{equation} \label{eq:Hankel}
	\mathcal{I}_n(k) =  \int_{0}^{R} J_n(k r) r \d r ,
	\end{equation}
	with the $n$-th order Bessel function of the first kind, $J_n(k r)$. In addition, we introduced shifted frequency coordinates 
	\begin{align}
	k^{\mp} = \sqrt{(k_x \mp k_0)^2 +k_y^2  }~~~\mathrm{and}~~~\xi^{\mp} = \arctan \left( \frac{k_y}{k_x \mp k_0} \right),
	\end{align}
	originating from the Fourier shift theorem
	\begin{equation}
	e^{\pm i k_0 x} f(\vec{r}) ~\overset{\text{FT}}{\laplace}~ \tilde{f}(k_x \mp k_0,k_y).
	\end{equation}
	\begin{figure}[t]
		\centering
		\includegraphics[width=\textwidth]{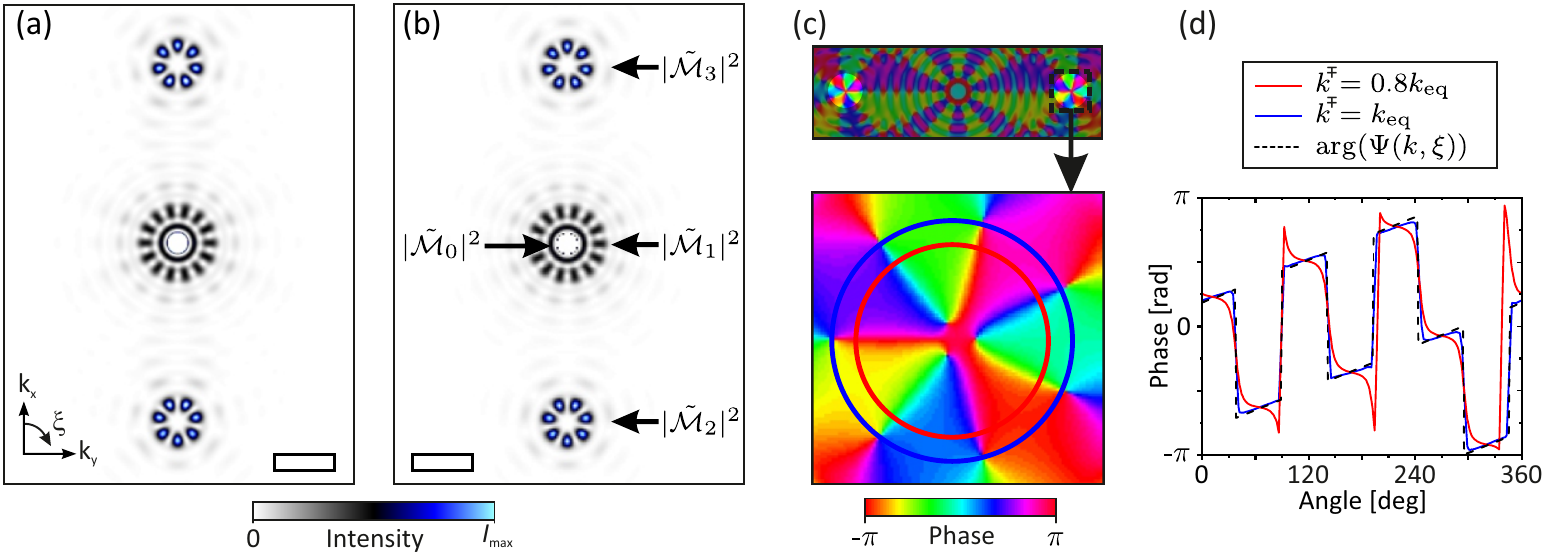}
		\caption{Accuracy of the holographically generated electron wave field with respect to the target wave. (a,b) Calculated electron diffraction pattern with (a) and without (b) considering mixing terms (cf. Eqs.\eqref{eq:FFT Maske} and \eqref{eq:calc_patt}, $\SI{3.7}{\micro m}$ aperture diameter, no mask binarization applied, $k_0= \SI{30}{\micro m^{-1}}$, $({m=4,~n=3})$, scale bar: $ \SI{10}{\micro m^{-1}}$). The targeted vortex structures with 7-fold rotational symmetry are described by $\vert \tilde{\mathcal{M}}_{2/3} \vert^2 $. (c) Phase distribution map of (a) in the far-field and zoom-in for the first diffraction order. Equation~\eqref{eq:M2,3} factorizes for $k^{\mp} \approx k_\mathrm{eq}$ around the first diffraction pattern, resulting in an approximate phase behaviour as contained in Eq.\eqref{eq:superpos}. (d) Wave function phase along circular paths marked in (c) for $k^{\mp}=k_\mathrm{eq}$ (blue) and $k^{\mp}=0.8 \, k_\mathrm{eq}$ (red), showing a step-like behaviour following the analytical expression as derived from Eq.\eqref{eq:superpos} (black dashed).}
		\label{fig:calc_diff_patt}
	\end{figure}
	For an approximation of the diffraction pattern, we neglect in the following the mixing terms $\tilde{\mathcal{M}}_j \tilde{\mathcal{M}}_k^*$, i.e.
	\begin{align}\label{eq:calc_patt}
	\Gamma = \vert \mathcal{F}\lbrace \mathcal{M} \rbrace \vert^2 = \sum_{j=0}^{3} \vert \tilde{\mathcal{M}}_j \vert^2  + \sum_{\substack{j,k = 0 \\ j \neq k}}^{3} \tilde{\mathcal{M}}_j \tilde{\mathcal{M}}_k^* \approx \sum_{j=0}^{3} \vert \tilde{\mathcal{M}}_j \vert^2.
	\end{align}
	This approximation is valid in the limit of sufficiently large $k_0$, compared to the widths of the Fourier-transforms of the apertured real-space target wave and its autocorrelation function $\tilde{\mathcal{M}_0}+\tilde{\mathcal{M}_1}$. In figure~\ref{fig:calc_diff_patt}, the calculated diffraction pattern is shown with and without considering the mixing terms. 
	For the chosen parameters, a weak interference between $\tilde{\mathcal{M}}_1$ and $\tilde{\mathcal{M}}_{2,3}$ is visible, resulting in a small distortion of the target wave formed in the side lobes.
	The main contributions to the diffraction pattern amount to
	\begin{align}
	&\vert \tilde{\mathcal{M}_0} \vert^2 = 36 \pi^2 \mathcal{I}^2_0(k) = 36\pi^2 \frac{R^2}{k^2} J_1^2(kR), \\
	&\vert \tilde{\mathcal{M}_1} \vert^2 = 8 \pi^2 \mathcal{I}^2_{(n+m)}(k)\bigg(1+ \cos\big[2(n+m)\xi + 2\kappa_\mathrm{SPM} \big] \bigg),\\
	&\vert \tilde{\mathcal{M}}_{2,3} \vert^2 = 4\pi^2 \bigg(  \mathcal{I}^2_n(k^{\mp})   + \mathcal{I}^2_m(k^{\mp})  \notag + 2 \mathcal{I}_n(k^{\mp}) \mathcal{I}_m(k^{\mp}) \\ & \quad~~~~~~~~~~~~~~~~~~~~~~ \times    \cos \left[ (n+m)\xi^{\mp} + \kappa_\mathrm{SPM} \pm (m-n) \frac{\pi}{2}    \right] \bigg). \label{eq:M2,3}
	\end{align}
	The first term $\vert \tilde{\mathcal{M}_0} \vert^2$ describes the zeroth diffraction order, surrounded by a structure with $c_{2(n+m)}$ rotationally symmetric density, given by $\vert \tilde{\mathcal{M}_1} \vert^2 $. The targeted mixed-OAM states with $c_{n+m}$-rotational symmetry occur in the first diffraction orders and are described by the terms of $\Gamma^{(\mp1)}(k,\xi) \equiv \vert \tilde{\mathcal{M}}_{2,3} \vert^2$, shifted by $\mp k_0$ in the $k_x$-direction.\\
	Notably, the Fourier transform of the apertured real-space target-wave contains a superposition of two OAM states, with the same topological charges as in the real-space target function. However, their relative amplitudes change with the reciprocal space distance $k^{\mp}$, due to the different $k^{\mp}$-dependence of the functions $\mathcal{I}_{n,m}(k^{\mp})$. For $m=4$ and $n=3$, both functions are equal for $k_\mathrm{eq}= \SI{3.43}{\micro m^{-1}}$ ($R= \SI{1.85}{\micro m}$), so that at this specific reciprocal space radius, the phase behavior of the Fourier-transformed target wave follows exactly Eq.\eqref{eq:superpos} with $\beta_0=1$ (cf. figure~\ref{fig:calc_diff_patt}(c,d)). \\	
	A more pronounced agreement with the target momentum wave function (cf. Eq.\eqref{eq:superpos}) and the holographically generated side lobes can be achieved by implementing different radially-dependent functions $f_q(r)$ (cf. Eq.\eqref{eq:decomposition_fourier}) in the construction of the holographic mask. In particular, a specific radial dependence $G(k)$ can be achieved by choosing the normalized (inverse) Hankel transform of $q$-th order $\mathcal{H}_q$ for the angular decomposition terms $f_q$. We employ this concept for the generation of a superposition state with a $k$-dependent relative phase $e^{iCk^2}$ between the OAM components combined with a Gaussian radial dependence $e^{-\frac{1}{2}C_2k^2}$, shown in figure~\ref{fig:Radial_Phase}. Here, the holographic mask is calculated as
	\begin{align}
	\mathcal{M}(r,\phi) =  \left|  e^{4i \phi} \cdot \mathcal{H}_4 \big\lbrace e^{iCk^2} e^{-\frac{1}{2}C_2k^2} \big\rbrace + e^{-3i \phi} \cdot \mathcal{H}_{3} \big\lbrace e^{-\frac{1}{2}C_2k^2}\big\rbrace    + e^{ik_0 x}\right|^2
	\end{align}
	for $m=4$ and $n=3$. Note that the imprinted phase singularity in the diffraction side lobes together with the employed mask binarization results in a vanishing probability density around ${k^{\mp}=0}$. Furthermore the thresholding inherent in the binarization scheme results in a finite mask extension.

	\section{Laser electric field and photoelectron momentum distribution in the MPI approach} 
	\label{app:electric_field} \label{app:MPI}
	
	In this section, we discuss the electric field (symmetry) properties and quantum dynamics of MPI on sodium atoms for mixed-OAM states. 
	The electric field for CRCP pulse sequences to generate $n$ vs. $m$ electron mixed-OAM states is given by \cite{Diels:2012:V,Wollenhaupt:2015:63}
	\begin{align}
	\vec{E}^- (t) &= \vec{E}^-_\text{r}(t) + \vec{E}^-_\text{b}(t)\notag \\
	&= \vec{e}_{1} \mathcal{E}_\text{r}(t) e^{-i  (\omega_\text{r} t + \varphi_\text{r} + \varphi_{\text{ce}}) }  + \vec{e}_{-1} \mathcal{E}_{\text{b}}(t) e^{-i  (\omega_{\text{b}} t + \varphi_{\text{b}} + \varphi_{\text{ce}}) } \notag \\ 
	&= \vec{e}_{1} \mathcal{E}_\text{r}(t) e^{-i  (n\omega t + \varphi_\text{r} + \varphi_{\text{ce}}) }  + \vec{e}_{-1} \mathcal{E}_{\text{b}}(t) e^{-i  (m\omega t + \varphi_{\text{b}} + \varphi_{\text{ce}}) },
	\end{align}
	using the polarization vectors $\vec{e}_{1}=  \frac{1}{\sqrt{2}} \begin{pmatrix}
	1 \\ 
	i
	\end{pmatrix}  $
	for LCP and $\vec{e}_{-1}=  \frac{1}{\sqrt{2}} \begin{pmatrix}
	1 \\ 
	-i
	\end{pmatrix}  $ for RCP, the respective field amplitudes $\mathcal{E}_{\text{r/b}}(t)$, the relative phases $\varphi_{\text{r/b}}$, the carrier-envelope phase $\varphi_{\text{ce}}$ and the frequencies $\omega = \frac{\omega_{\text{b}}}{m} = \frac{\omega_{\text{r}}}{n}$. For equal envelopes $ \mathcal{E}_{\text{b}}(t) = \mathcal{E}_{\text{r}}(t) \equiv \mathcal{E}_0(t)$ and without the phases we find
	\begin{align}
	\vec{E}^-(t) =  \mathcal{E}_0(t) \left(\vec{e}_{1} e^{-i n \omega t } +  \vec{e}_{-1}e^{-i m \omega t } \right). 
	\end{align}
	Therefore the real-valued laser electric field is given by
	\begin{align}
	\vec{E}(t) &=\Re \lbrace \vec{E}^-(t) \rbrace = \frac{1}{\sqrt{2}} \mathcal{E}_0(t)   \begin{pmatrix}
	\cos \left( n \omega t \right) + \cos \left( m \omega t \right)\\ 
	\sin \left( n \omega t \right)-\sin \left( m \omega t \right)
	\end{pmatrix}\notag \\
	&=\sqrt{2} \mathcal{E}_0(t)  \cos \left( \frac{(n+m)}{2} \omega t \right)  \begin{pmatrix}
	\cos \left( \frac{(n-m)}{2} \omega t \right)\\ 
	\sin \left( \frac{(n-m)}{2} \omega t \right)
	\end{pmatrix},
	\end{align}
	using $\Phi = \arctan \left( \frac{E_y}{E_x} \right) = \frac{n-m}{2} \omega t$ leads to an azimuthal velocity $\dot{ \Phi} = \frac{n-m}{2} \omega $. Note that the MPI with the \textit{blue} field component ($\omega_{\text{b}} = m \omega$) corresponds to a photonicity of $N_{\text{p}}^{\text{blue}} = n$, while the one for the \textit{red} component  ($\omega_{\text{r}} = n \omega$) leads to $N_{\text{p}}^{\text{red}} = m$ to ensure interband interferences. Hence, the excitation with an \bichrom{n}{m} field leads to $m$ vs. $n$ photon processes \cite{Kerbstadt:2019:NC:658}. For this reason we rewrite the azimuthal velocity 
	\begin{equation}\label{eq:angular_velocity_field}
	\omega_{ \Phi}(m,n) =  \frac{m-n}{2} \omega ,
	\end{equation}
	which is shown to be constant within such a propeller-type polarization profile. Here $n$ and $m$ now represent the respective photonicities $N_\text{p}$. This can be associated with an induced azimuthal probability current $\vec{j}$ in the resulting photoelectron wave packet (cf. \ref{app:topo_charge}).\\
	A rotation of our bicircular field around an angle $\zeta$, i.e. using a $\lambda/2$-waveplate under $\zeta/2$, is represented by
	\begin{equation}\label{eq:rot_lambda}
	\mathbf{R}(\zeta) (\vec{e}_1 \mathcal{E}_{\text{r}} + \vec{e}_{-1} \mathcal{E}_{\text{b}}) = \vec{e}_1 \mathcal{E}_{\text{r}} e^{-i \zeta} + \vec{e}_{-1} \mathcal{E}_{\text{b}} e^{i \zeta},
	\end{equation}
	with $\mathbf{R}(\zeta)$ as the active rotation matrix in mathematical positive direction and the respective polarization vectors $\vec{e}_{\pm 1} $.\\
	In the following we derive that these polarization-shaped bichromatic fields with commensurable frequencies enable, via preselected $\sigma^{\pm}$-transitions (\emph{spectral double slit}), the generation of OAM states described by Eq.\eqref{eq:superpos}. The perturbative description of the MPI (with $N_{\text{p}}$ photons) of sodium atoms leads to~Eq.\eqref{EQ8} in momentum spherical coordinates $(k,\xi,\vartheta)$ \cite{Kerbstadt:2020:43,Kerbstadt:2019:APX:1672583,Kerbstadt:2019:NC:658}.
	The factor $i^{N_{\text{p}}}$ represents the phase from $N_{\text{p}}$-th order perturbation theory \cite{Meshulach:1999:PRA:1287,Dudovich:2005:PRL:083002,Dudovich:2001:PRL:47}. Together with optical phases and the rotation angle $\zeta$ of a $\lambda/2$-waveplate (cf. Eq.\eqref{eq:rot_lambda}) we find for $\sigma^{\pm}$-transitions with $N_{\text{p}}$ photons
	\begin{align}
	\tilde{\psi}_{l,\pm m}(k,\xi,\vartheta) &=  \psi_{l,\pm m} (k,\xi,\vartheta) e^{-i N_{\text{p}} (\varphi_{{\text{r/b}}} + \varphi_{{\text{ce}}} )} e^{\pm i N_{\text{p}} \zeta}.
	\end{align}
	Since solely $\sigma^{\pm}$-transitions are discussed (cf. figure~\ref{fig:Spatial_Modulation_Results}(b)) we denote $N_{\text{p}} \equiv m$.
	This leads to the photoelectron wave function of the superposition state
	\begin{align}
	\Psi_{\text{MPI}}(k,\xi,\vartheta) &= \tilde{\psi}_{m, m}(k,\xi,\vartheta) + \tilde{\psi}_{n, - n}(k,\xi,\vartheta) \notag \\
	&\propto  \psi_{m, m}(k,\xi,\vartheta) e^{i \kappa_\mathrm{MPI}} + \psi_{n, -n}(k,\xi,\vartheta)  ,
	\end{align}
	with 
	\begin{equation}
	\kappa_\mathrm{MPI} =  - [m \varphi_{\text{r}} -n \varphi_{\text{b}} + (m-n) \varphi_{{\text{ce}}} - (m+n)\zeta].
	\end{equation}
	Further simplification is achieved by the approximations $\mathcal{P}_{l+1, l+1} \approx (-1) \mathcal{P}_{l, l} $ for the associated Legendre polynomials around $\vartheta = \frac{\pi}{2}$ and the radial part $\mathcal{R}_{l+1} \approx \mathcal{R}_{l} $ assuming Gaussian envelopes, leading to
	\begin{align}
	\Psi_{\text{MPI}}(k,\xi,\vartheta) &\approx~ \mathcal{R}_n(k) \mathcal{P}_{n,n}[\cos(\vartheta)]  \bigg( i^m (-1)^{m-n} e^{i m \xi} e^{i \kappa_\mathrm{MPI}} + i^n (-1)^n  e^{-in \xi} \bigg) \notag \\
	& \propto \mathcal{R}_n(k) \mathcal{P}_{n,n}[\cos(\vartheta)]  \bigg(e^{i m \xi} e^{i \gamma} + e^{-in \xi} \bigg),  
	\end{align}
	and identifying $\gamma = \kappa_\mathrm{MPI} + (m-n) \frac{\pi}{2} + m \pi $. Within the approximation of $\vartheta \approx \frac{\pi}{2}$, introduced above, we find
	\begin{equation}
	\Psi_{\text{MPI}}(k,\xi,\vartheta) \approx G_{\text{MPI}}(k) \bigg(e^{i m \xi} e^{i \gamma} + e^{-in \xi} \bigg), 
	\end{equation}
	with $G_{\text{MPI}}(k) = \mathcal{R}_n(k) \mathcal{P}_{n,n}[\cos(\vartheta)]$ and therefore an analogous description to Eq.\eqref{eq:superpos}. The resulting electron density is given by
	\begin{equation}\label{app:eq:electron_density_MPI}
	\vert \Psi_{\text{MPI}} \vert^2 \approx  \vert G_{\text{MPI}}(k) \vert^2 \bigg[ 1 + \cos\bigg( \left[ m+n \right] \xi + \gamma   \bigg)  \bigg],
	\end{equation}
	leading to Eq.\eqref{eq:superpositionState}.
	
	\section{Quantum mechanical properties of mixed-OAM states}
	\label{app:topo_charge}
	
	The topological charge along with the probability current $\vec{j}$ and the expectation value of the $z$-component of the orbital angular momentum $\langle L_z \rangle$ are key quantities describing the properties of the OAM superposition states realized in both experiments. The probability current \cite{Cohen-Tannoudji:1991}
	\begin{equation}
	\vec{j} = \frac{\hbar}{m} \Im 	\left[\Psi^* \vec{\nabla} \Psi\right] =  \frac{\hbar}{m} \rho \cdot  \vec{\nabla} \arg(\Psi) \propto \rho \vec{\omega}
	\end{equation}
	describes the flow of probability-density $\rho$ in the system. The angular frequencies $\vec{\omega} \propto \vec{\nabla} \arg(\Psi)$ are associated with the angular group velocity of the wave function. Hence, we find for Eq.\eqref{eq:superpos}
	\begin{align}
	\vec{j} &   \propto  \vert \Psi \vert^2 \frac{(m - n)}{2 k} \vec{e}_{\xi} + \mathcal{O}(\beta_0 -1)  \approx \rho \frac{m-n}{2} \vec{e}_{\xi},
	\end{align}
	with the azimuthal group velocity $\omega_{\xi} \propto \frac{m-n}{2}$. Note that this velocity corresponds to the angular velocity of the laser electric field in Eq.\eqref{eq:angular_velocity_field}, i.e.,
	\begin{equation}
	\omega_{\xi}(m,n) \propto \omega_{\Phi}(m,n).
	\end{equation}
	A further analogy can be drawn by the investigation of the expectation value of the angular momentum in $z$-direction $\langle L_z \rangle$. For the general superposition state in Eq.\eqref{eq:superpos} we find
	\begin{equation}
	\langle L_z \rangle = \langle \Psi \vert L_z \vert \Psi \rangle \propto \frac{\beta_0^2 m -n}{1 + \beta_0^2} \approx \frac{m-n}{2} + \mathcal{O}(\beta_0-1).
	\end{equation}
	This result is in agreement with the integrated probability current in azimuthal direction, given by
	\begin{equation}
	\int_{0}^{2 \pi} (\vec{j})_{\xi} \text{d}\xi \propto \frac{\beta_0^2 m -n}{1 + \beta_0^2} \approx \frac{m-n}{2} + \mathcal{O}(\beta_0-1)
	\end{equation}
	and representing an azimuthal velocity of the electron density. Another quantity which is commonly discussed in this context is the topological charge $\ell$, describing the accumulated phase following a contour $\mathcal{C}$ and defined as \cite{Bliokh:2017:PR:1} 
	\begin{equation}\label{eq:topo_charge}
	2 \pi \ell = \oint_{\mathcal{C}} \vec{\nabla} \arg(\Psi) \cdot \text{d}  \vec{k},
	\end{equation}
	where $\mathcal{C}$ is a contour enclosing the phase singularity. Combining Eq.\eqref{eq:superpos} with Eq.\eqref{eq:topo_charge} using $\beta_0 \in \mathbb{R}_+$ we get
	\begin{align}
	\vec{\nabla} \arg(\Psi) = \frac{\beta_0(m-n)\cos((m+n)\xi)-\beta_0^2 n + m}{(\beta_0^2 + 2 \beta_0 \cos((m+n)\xi) + 1)k \sin(\vartheta)}
	\end{align}
	\begin{figure}[t]
		\centering
		\includegraphics[width=\textwidth]{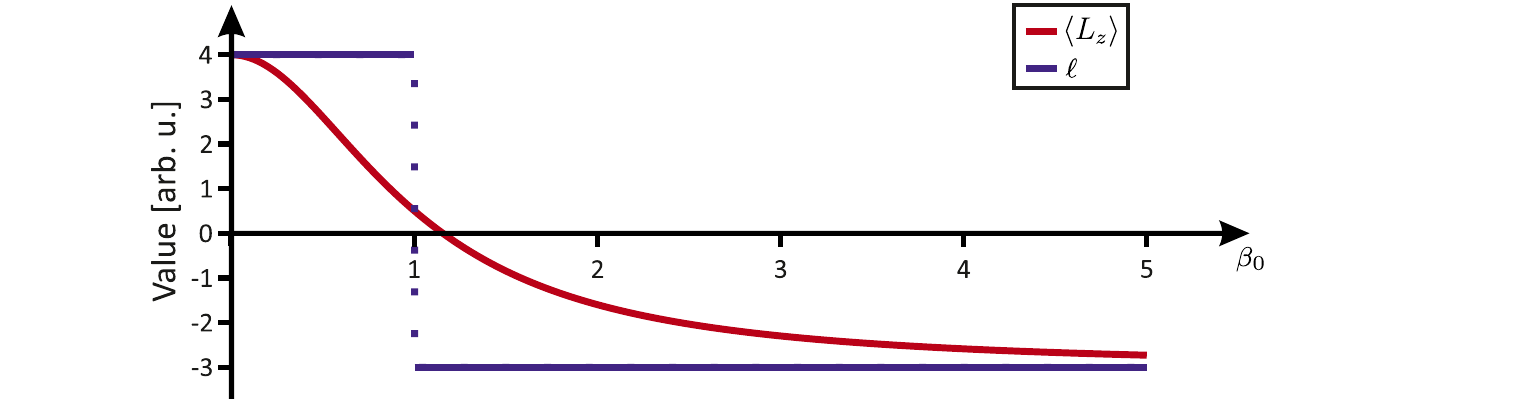}
		\caption{Calculated expectation value of the $z$ component of the angular momentum $\langle L_z \rangle $ (red) and the topological charge $\ell$ as a function of the amplitude factor $\beta_0$ (violet) for a OAM superposition state with $m=4$ and $n = 3$.}
		\label{fig:App_Current}
	\end{figure}
	and find
	\begin{align}
	\ell  &= \frac{1}{2 \pi} \int_0^{2\pi} \nabla \arg(\Psi ) \d \xi  = 	\frac{1}{2 \pi} \int_0^{2\pi}  \frac{(m-n)}{2} \frac{ \frac{m-\beta_0^2 n}{\beta_0 (m-n)} + \cos([m+n]\xi) }{ \frac{1+\beta_0^2}{2\beta_0} + \cos([m+n]\xi ) }  \d \xi.
	\end{align}
	With $(m+n ) \in \mathbb{N}$ it can be written as
	\begin{align}
	\ell &= \frac{1}{2} \left(  \text{sgn}\left[1-\beta_0^2\right](m+n) - (n-m) \right) = 
	\begin{cases}
	m &;~ \beta_0 <1 \\
	\frac{m-n}{2} &;~ \beta_0 = 1 \\
	-n &;~ \beta_0 >1.
	\end{cases}
	\end{align}
	This piecewise behaviour differs from the ones of $\langle L_z \rangle$ or $\vec{j}$ and is depicted in figure~\ref{fig:App_Current} exemplarily for  $m=4$ and $n = 3$.
	
	\newpage

	\section*{References}
	\bibliographystyle{ieeetr}

\end{document}